\documentclass[aps,prl,preprint,superscriptaddress]{revtex4-1}
\usepackage{graphics}      % standard graphics specifications
\usepackage[english]{babel}
\usepackage{graphicx}      % alternative graphics specifications
\usepackage{longtable}     % helps with long table options
\usepackage{url}           % for on-line citations
\usepackage{bm}            % special 'bold-math' package
\usepackage{amsmath}
\usepackage{textcomp}
\usepackage{amsfonts}
\usepackage{amssymb}
\usepackage{float}
\usepackage{hhline}
\usepackage{braket}
\usepackage{lineno}
\begin{document}

\mathchardef\mhyphen="2D

\title{\textit{First-principles} treatment of Mott insulators: linearized QSGW+DMFT approach}

\author{Sangkook Choi}
% \email[]{Your e-mail address}
% \homepage[]{Your web page}
% \thanks{}
% \altaffiliation{}
\affiliation{Condensed Matter Physics and Materials Science Department, Brookhaven National Laboratory, Upton, NY, USA}
\affiliation{Department of Physics and Astronomy, Rutgers University, Piscataway, New Jersey 08854, USA}
\author{Andrey Kutepov}
% \email[]{Your e-mail address}
% \homepage[]{Your web page}
% \thanks{}
% \altaffiliation{}
\affiliation{Department of Physics and Astronomy, Rutgers University, Piscataway, New Jersey 08854, USA}
\author{Kristjan Haule}
% \email[]{Your e-mail address}
% \homepage[]{Your web page}
% \thanks{}
% \altaffiliation{}
\affiliation{Department of Physics and Astronomy, Rutgers University, Piscataway, New Jersey 08854, USA}
\author{Mark van Schilfgaarde}
% \email[]{Your e-mail address}
% \homepage[]{Your web page}
% \thanks{}
% \altaffiliation{}
\affiliation{Department of Physics, Kings College London, Strand, London WC2R 2LS, United Kingdom}
\author{Gabriel Kotliar}
\email[]{Correspondence and requests for materials should be addressed to G.K. (kotliar@physics.rutgers.edu)}
% \homepage[]{Your web page}
\affiliation{Condensed Matter Physics and Materials Science Department, Brookhaven National Laboratory, Upton, NY, USA}
\affiliation{Department of Physics and Astronomy, Rutgers University, Piscataway, New Jersey 08854, USA}

% Collaboration name if desired (requires use of superscriptaddress
% option in \documentclass). \noaffiliation is required (may also be
% used with the \author command).
% \collaboration can be followed by \email, \homepage, \thanks as well.
% \collaboration{}
% \noaffiliation

\date{\today}

% \pacs{71.15.-m}{Methods of electronic structure calculations}
% \pacs{71.27.+a}{Strongly correlated electron systems; heavy fermions}
% \pacs{74.72.Cj}{Insulating parent compounds}

\begin{abstract}
The theoretical understanding of emergent phenomena in quantum materials is one of the greatest challenges in condensed matter physics. In contrast to simple materials such as noble metals and semiconductors, macroscopic properties of quantum materials cannot be predicted by the properties of individual electrons. One of the examples of scientific importance is strongly correlated electron system. Neither localized nor itinerant behaviors of electrons in partially-filled $3d$, $4f$, and $5f$ orbitals give rise to rich physics such as Mott insulators, high-temperature superconductors, and superior thermoelectricity, but hinder quantitative understanding of low-lying excitation spectrum. Here, we present a new \textit{first-principles} approach to strongly correlated solids. It is based on a combination of the quasiparticle self-consistent GW approximation and the Dynamical Mean Field Theory (DMFT). The sole input in this method is the projector to the set of correlated orbitals for which all local Feynman graphs are being evaluated. For that purpose, we choose very localized quasiatomic orbitals spanning large energy window, which contains most strongly-hybridized bands as well as upper and lower Hubbard bands. The self-consistency is carried out on the Matsubara axis. This method enables the \textit{first-principles} study of Mott insulators in both their paramagnetic (PM) and antiferromagnetic (AFM) phases. We illustrate the method  on the archetypical charge transfer correlated insulators La$_2$CuO$_4$ and NiO, and obtain spectral properties and magnetic moments in good agreement with experiments.
%   We present a new \textit{first principles} approach to strongly correlated
%   solids. It is based on a combination of the quasiparticle self-consistent GW approximation and the Dynamical Mean Field Theory
%   (DMFT). The sole input in this method is the projector to the set of
%   correlated orbitals for which all local Feynman graphs are being
%   evaluated. For that purpose we choose very localized quasiatomic
%   orbitals spanning large energy window, which contains most strongly-hybridized bands as well as upper and lower Hubbard bands. The self-consistency is carried out on the Matsubara axis.
% %
%   This method enables the first principles study of Mott insulators in
%   both their paramagnetic (PM) and antiferromagnetic (AFM) phase.  
%   We illustrate the method  on the archetypical charge transfer correlated
%   insulator La$_2$CuO$_4$, and obtain spectral properties and magnetic moments in 
% good agreement with experiments.
\end{abstract}

\maketitle

% \preprint{}

% Title of paper



\textit{Introduction}. The \textit{first-principles} description of
strongly-correlated materials is currently regarded as one of the
greatest challenges in condensed matter physics. The interplay between
correlated electrons in open $d$- or $f$- shell and itinerant band
states gives rise to rich physics that makes these materials
attractive for a wide range of applications such as oxide electronics,
high-temperature superconductors and spintronic devices. Various
theoretical approaches are currently being pursued \cite{anisimov_strong_2000}. 
One of the most
successful approaches is the dynamical mean field theory (DMFT)
\cite{georges_dynamical_1996}. In combination with density functional
theory \cite{anisimov_first-principles_1997,lichtenstein_textbf_1998},
it has described many features of strongly-correlated materials
successfully and highlighted the surprising accuracy of treating
correlations local to a small subset of orbitals exactly, while
treating the reminder of the problem in a static mean field
manner.\cite{kotliar_electronic_2006}.

% Correlations are built into the full electron Green's function ($G$), through electron self-energy ($\Sigma$)
% \begin{equation}
% G(\mathbf{r},\mathbf{r}',i\omega_n)= 
% \bra{\mathbf{r}} \left(i\omega_n\mathbf{1}-\hat{H}_H-\hat{\Sigma}(i\omega_n)\right)^{-1}   \ket{\mathbf{r}'} 
% \label{eq:GreenF}
% \end{equation}
% where $\mathbf{r}$ is position vector, $\omega_n$ is Matsubara frequency and $\hat{H}_H$ is the Hartree Hamiltonian.
% The successes of the DMFT method suggest that in certain energy range
% a good expression for the self-energy $\Sigma$ is provided by
% \begin{equation}
% \Sigma(\mathbf{r},\mathbf{r}',i\omega_n)= \sum_{\mathbf{R} \alpha \beta} P^*_{\mathbf{R} \alpha}(\vr) \Sigma^{\alpha,\beta}_{loc}(i\omega_n) P_{\mathbf{R} \beta}(\vr')  + V(\vr,\vr') \label{eq:local_ansatz}
% \end{equation}
% with  $\Sigma_{loc}$ described by an impurity model, where $\mathbf{R}$ is
% lattice vector and $P_{\mathbf{R} \alpha}(\vr)$ is the projector to a set of local orbitals
% $\alpha$ centered at $\mathbf{R}$.

The numerous successes of DMFT in different classes of correlated materials revived the interest in the long-sought goal of achieving a diagrammatically controlled approach to the quantum many-body problem  of  solids, starting from the Green's function $G$ and the screened Coulomb interactions $W$ ~\cite{almbladh_variational_1999,chitra_effective-action_2001}.  The lowest order diagram in perturbation theory in this functional gives rise to the GW approximation \cite{hedin_new_1965} while the local approximation applied to the most correlated orbitals gives
rise to an  extended DMFT approach to the electronic structure problem \cite{chitra_effective-action_2001}. The addition of the GW and DMFT graphs was proposed and implemented in model Hamiltonian studies \cite{sun_extended_2002} and in realistic electronic structure \cite{biermann_first-principles_2003,kotliar_model_2001}.
 There is now  intense activity in this area with many recent publications
\cite{tomczak_combined_2012,sakuma_electronic_2013,taranto_comparing_2013,tomczak_asymmetry_2014} triggered by advances in the quality of the impurity solvers \cite{werner_continuous-time_2006,haule_quantum_2007}, insights into the analytic form of the high-frequency behavior of the self-energy \cite{casula_dynamical_2012} and improved electronic
structure codes. 

Several conceptual issues remain to be clarified before the long
sought goal of a robust electronic structure method for solids is
attained. The first issue 
is the choice of 
local orbitals % ($P_{R\alpha}(\vr)$ in Eq.~\eqref{eq:local_ansatz})
on which to
perform the DMFT method (summation of all local Feynman graphs).
The second issue is the level of self-consistency that should be used
in the calculation of various parts of the diagrams included in the
treatment (free or bare Green's function $G_0$ vs self-consistent
interacting Green's functions $G$). These central issues are addressed in this letter. 

The self-consistency issue appears already at the lowest order, namely,
the GW level, and it has been debated over time. The corresponding
issue in GW+DMFT is expected to be at least as important, but has not
been explored, except for model
Hamiltonians~\cite{sun_many-body_2004,hansmann_long-range_2013}. At
the GW level, it is now well established that Hedin's fully self-consistent formulation~\cite{hedin_new_1965}, while producing good
total energies in solids  \cite{kutepov_ground-state_2009}, atoms and molecules \cite{stan_fully_2006,stan_levels_2009},
does not produce a good approximation to the spectra of
even 3D electron gas and aluminum in comparison to non-self-consistent GW results \cite{holm_fully_1998,kutepov_ground-state_2009}. Instead, using
a free (quasiparticle) Green's function in the evaluation of the
polarization graph of the GW method gives much better results for
spectral functions. This is the basis of the one-shot quasiparticle (QP)
GW, starting from LDA \cite{hybertsen_electron_1986} or from others \cite{rinke_combining_2005}. Unfortunately, the answer
depends on the starting point. A solution for this problem is to
impose a self-consistency equation to determine $G_0$.
This method, called the quasiparticle self-consistent GW (QSGW) \cite{kotani_quasiparticle_2007}, is very successful  reproducing the  spectra of many systems
\cite{kotani_quasiparticle_2007}. 
How to combine it with DMFT is an important open challenge \cite{tomczak_many-body_2012, tomczak_qsgw+dmft:_2014}. 

\begin{figure}[t]
  \centering
  \includegraphics[width=0.70\columnwidth]{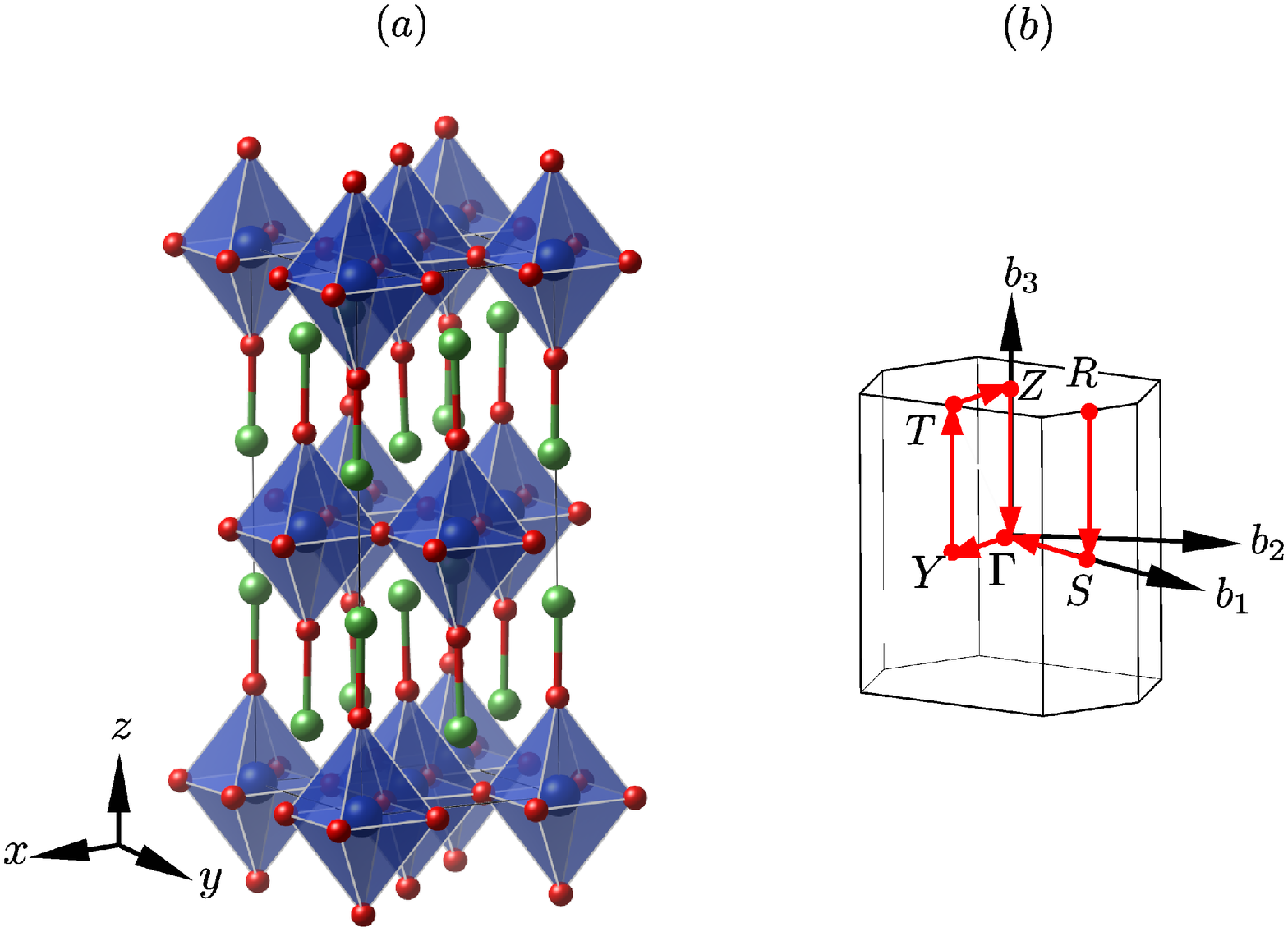}
  \caption {Atomic structure and first Brillouin Zone of La$_2$CuO$_4$. (a) Atomic structure of La$_2$CuO$_4$ in the single face-centered orthorhombic phase. Lanthanum atoms are represented by green spheres, copper atoms by blue spheres in the blue octahedrons, and oxygen atoms by red spheres. The structure is characterized by an alternating rotation of successive Cu$O_6$ octahedra along the $x$ direction. (b) First Brillouin zone of single face-centered orthorhombic phase. Red lines show the path along which electronic bandstructures are plotted in Fig. 2(c) and Fig. 3.}
\end{figure}

Previous GW+DMFT studies typically used a $G_0$ which depends on the LDA starting point, and projectors spanning a relatively small
energy window \cite{tomczak_combined_2012,sakuma_electronic_2013,taranto_comparing_2013,tomczak_asymmetry_2014}. 
In this work, we propose a different approach to the
level of self-consistency and the choice of the DMFT orbital. We do a self-consistent QSGW calculation and then calculate local self-energy using DMFT with static $U_d$ and $J_H$ without feedback to non-local self-energy within GW. For the DMFT step, we choose a very localized orbital spanning large energy window which contains most strongly-hybridized bands as well as upper and lower Hubbard bands.

\begin{figure}[t]
  \centering
  \includegraphics[width=0.70\columnwidth]{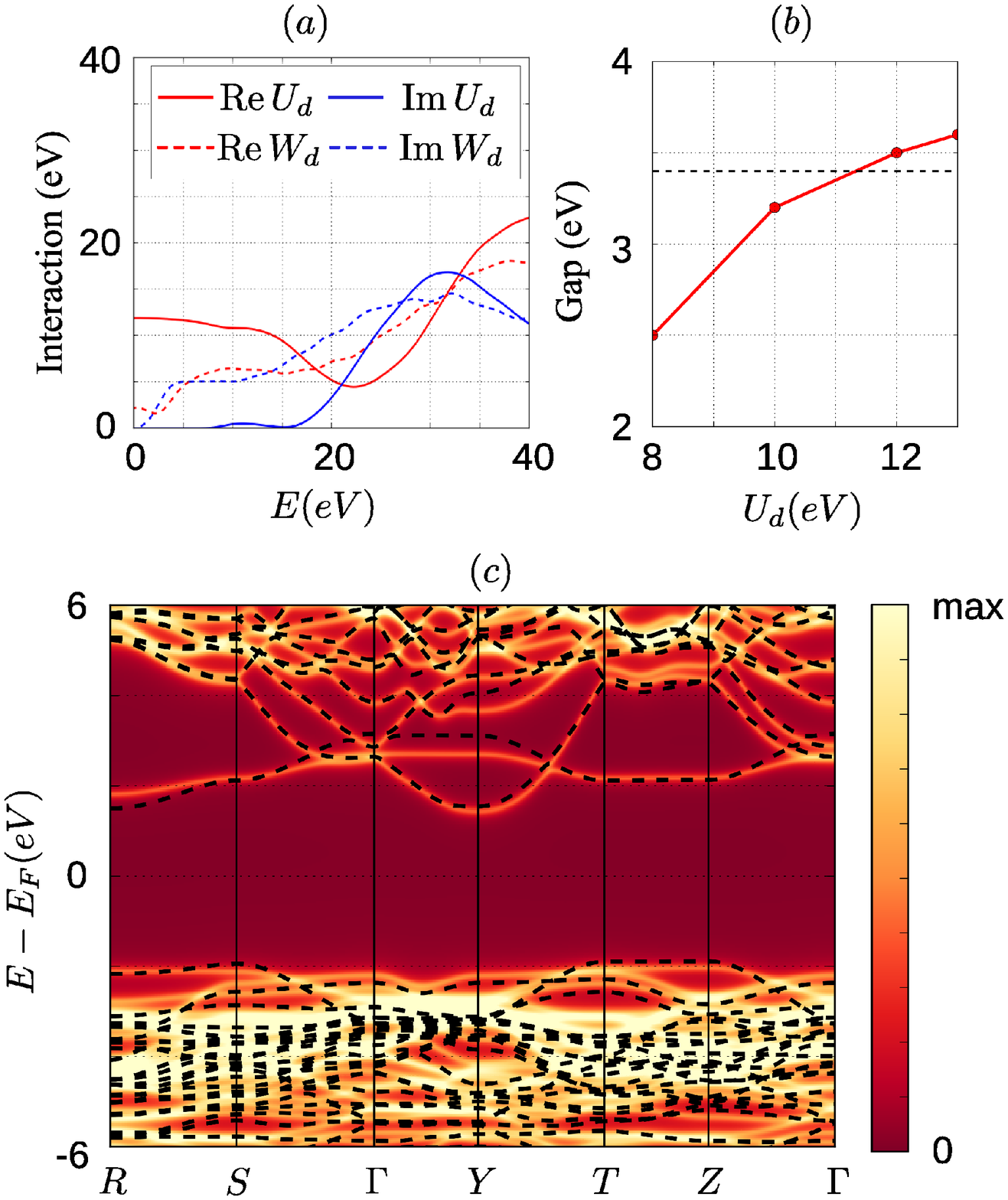}
  \caption {Hubbard U associated with Cu-3$d$ orbitals in La$_2$CuO$_4$. (a) Frequency dependence of $W_d$ (dashed lines) and $U_d$ (full lines) of La$_2$CuO$_4$ with a $\chi_{QP}^{low}$ defined in the energy window $E_F\pm 10eV$. Real and imaginary parts of the parameter are marked by red and blue colors, respectively. (b) Bandgap dependence on $U_d$, in La$_2$CuO$_4$, evaluated with impurity self-energy within spin-polarized GW approximation with $J_H$=1.4eV. The Black dashed line represents bandgap within spin-polarized MQSGW. (c) Spectral function of La$_2$CuO$_4$ with $U_d$=12eV and $J_H$=1.4eV. The black dashed-lines show bandstructures within spin-polarized MQSGW}
\end{figure}

In the LDA+DMFT context, the choice of very localized orbitals
has provided a great deal of universality since the interactions do
not vary much among compounds of the same family. This has been
demonstrated in the studies of iron pnictides \cite{yin_kinetic_2011} and
transition metal oxides \cite{haule_covalency_2014}. This choice results in a second 
advantage as we will show below, namely the frequency dependence of the
interaction matrix
can be safely ignored. Having chosen the correlated orbitals, all the other
parameters are self-consistently determined. This is the first
\textit{ab initio} quasiparticle self-consistent GW+DMFT implementation and 
the first study on a paramagnetic Mott insulator within the  GW+DMFT
method.

\begin{figure*}
  \centering
  \includegraphics[width=0.70\columnwidth]{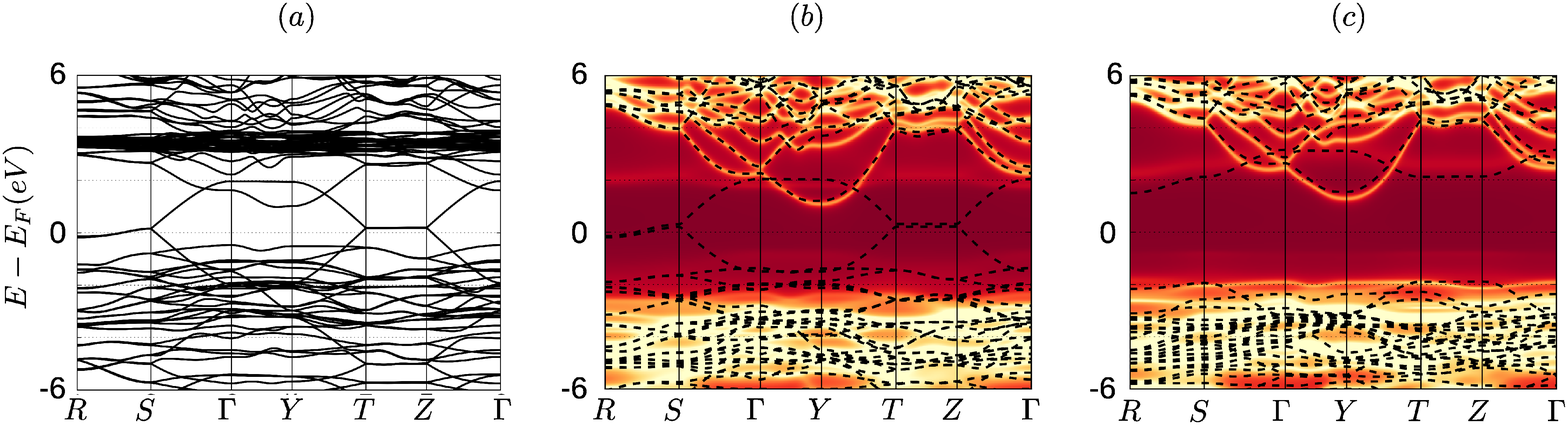}
  \caption {The low-energy spectral function of La$_2$CuO$_4$. (a) Electronic bandstructures of La$_2$CuO$_4$ within LSDA and spectral functions from (b) non-spin-polarized MQSGW+DMFT (c) and spin-polarized MQSGW+DMFT calculations along the path shown in Fig. 1(b). The Dashed lines in (b) and (c) represent electronic bandstructures within non-spin-polarized MQSGW and spin-polarized MQSGW, respectively}
\end{figure*}

\textit{Results}. Fig. 2(a) shows the frequency dependence of real and
imaginary parts of $U_d$ of La$_2$CuO$_4$ shown in Fig. 1. It is calculated on an imaginary frequency
axis and analytically continued by a maximum entropy method \cite{jarrell_bayesian_1996}. We also plot the fully screened Coulomb interaction $W_d$ for comparison. Static $U_d$ is 12.0 eV and $U_d$ remains almost
constant up to $10\,$eV. In contrast, in $W_d$, there are
several peaks due to low-energy collective excitations below $10\,$eV. At very high
energy, $U_d$ approaches the bare coulomb interaction of
$28\,$eV. Static value of $U_{pd}$ is 2.0 eV, much smaller than $U_d$, hence we don't discuss its treatment further\cite{optimal_u}. Calculated $J_H$ is $1.4\,$eV and has negligible frequency
dependence. By contrast, conventional constrained-RPA, in which 10 bands of mostly
Cu-$3d$ character are excluded from screening, results in static
$U_d=7.6\,$eV, which is too small to open the Mott gap, and which is
also inconsistent with photoemission experiments on CuO charge
transfer insulators \cite{ghijsen_resonant_1990}. 

We introduced a complementary method to compute the static $U_d$. The key idea is to first calculate the excitation spectra of La$_2$CuO$_4$ within MQSGW+DMFT using local GW (with a static $U_d$) as the impurity solver and then determine $U_d$, by finding the value that best matches the full spin-polarized MQSGW spectra. The procedure starts from the non-spin-polarized MQSGW band structure without magnetic long-range order. We then allow spontaneous magnetic long-range order by embedding a polarized impurity self-energy for the Cu-3$d$ electrons computed in a local GW approximation. We find that indeed magnetic ordering associated with Cu-3$d$ is captured by spin-polarized local MQSGW using a static value of $U_d$ and $J_H$ and spectral properties such as energy gap are very similar in value to the full spin-polarized MQSGW spectra. In Fig. 2(b), we allowed $U_d$ to vary between 8-13$\,$eV (at fixed $J_H=1.4\,$eV) and we plot the size of the indirect gap. The gap size of this method matches the gap of spin-polarized MQSGW when $U_d\approx 12\,$eV. If this choice of $U_d$ and $J_H$ is correct, the resulting spectra must be similar to the prediction of spin-polarized MQSGW method. We show this comparison in Fig. 2(c) to confirm a good match. In addition, the relative position of the Cu-$d$ band (the lowest energy conduction band at S) to the La-$d$ band (the lowest energy conduction band at Y) is also well matched justifying the approximation of $\hat{\Sigma}^{DC}(i\omega_n)\simeq\hat{\Sigma}^{DC}(i\omega_n=0)$. $\Sigma^{DC}(i\omega_n=0)$ for Cu-$d_{x^2-y^2}$ orbital differs from nominal double counting energy \cite{haule_dynamical_2010} by only $1\%$, highlighting again the advantages of using a broad window and narrow orbitals.

\begin{figure}[t]
  \centering
  \includegraphics[width=0.70\columnwidth]{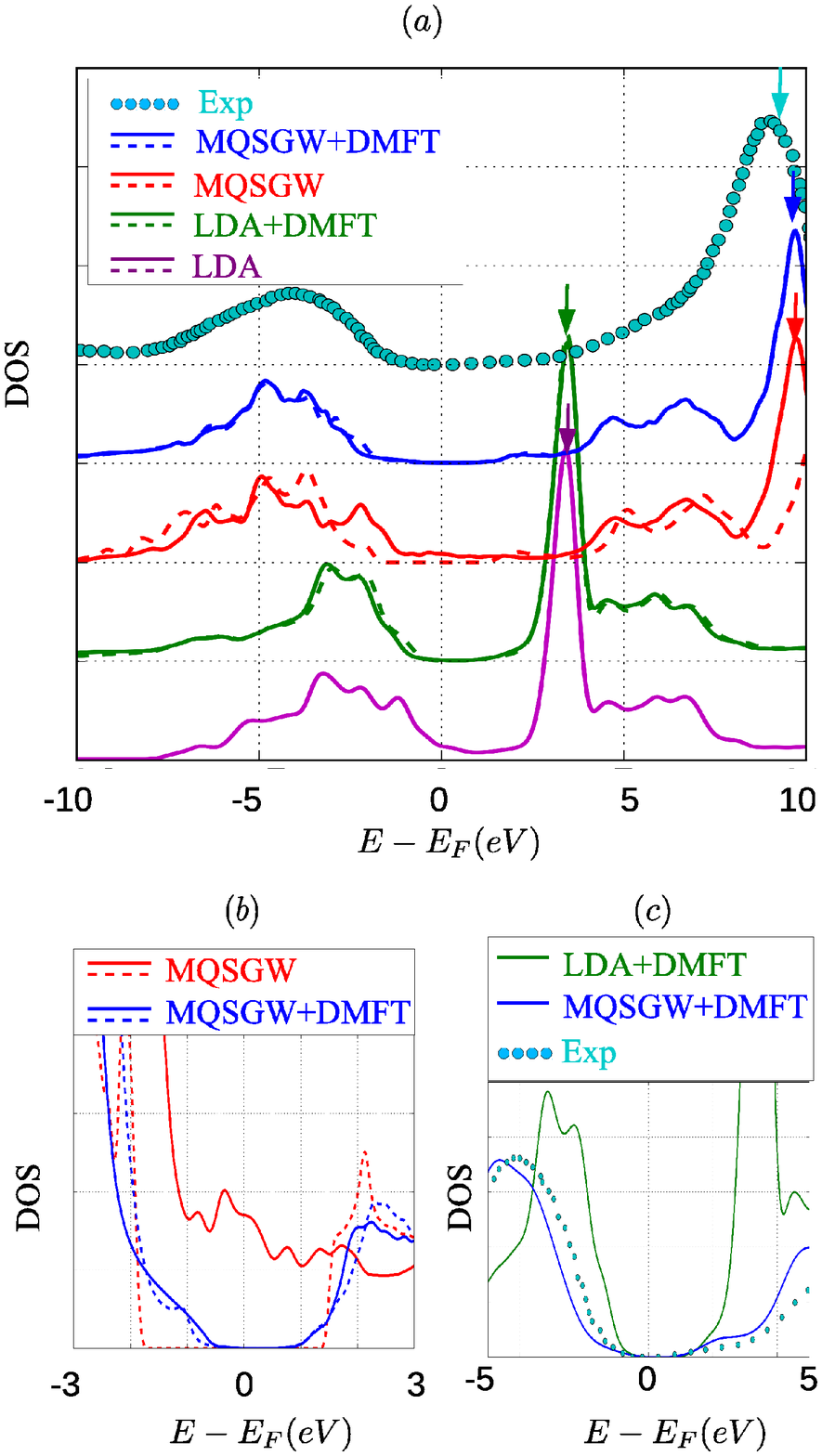}
  \caption {The density of states of La$_2$CuO$_4$. (a) Total density of states of La$_2$CuO$_4$ from LDA (magenta), LDA+DMFT(green), MQSGW (red), and MQSGW+DMFT (blue). Full lines and dashed-lines represent quantities within non-spin-polarized and spin-polarized versions of each calculation, respectively. The cyan dotted line shows photoemission/inverse-photoemission data \cite{nucker_experimental_1987}. The Positions of La-$f$ peaks are marked by arrows. (b) A zoom-in view in the low-energy region. (c) The overlap of total density of states of La$_2$CuO$_4$ within LDA+DMFT as well as MQSGW+DMFT and photoemission/inverse-photoemission data \cite{nucker_experimental_1987}}
\end{figure}

We now discuss the magnetic moment associated with Cu and the electronic excitation spectra of La$_2$CuO$_4$ by using MQSGW+DMFT (with $U_d=12.0eV$,
$J_H=1.4eV$) in which the impurity is solved by the numerically exact
CTQMC \cite{haule_quantum_2007,werner_continuous-time_2006} and compare them with other methods. LSDA does not have a magnetic solution. In contrast, spin-polarized MQSGW, QSGW \cite{kotani_quasiparticle_2007}, and MQSGW+DMFT predict 0.7 $\mu_B$, 0.7 $\mu_B$, and 0.8 $\mu_B$, respectively. This is consistent with experimental measurements, although the later span quite large range
$0.4\mu_B \sim 0.8\mu_B$~\cite{borsa_staggered_1995,reehuis_crystal_2006,vaknin_antiferromagnetism_1987}. %We find that MQSGW opens too large gap in the antiferromagnetic (AFM) phase, while it remains metallic in the paramagnetic (PM) phase. LDA+DMFT predict too small excitation energy of La-$f$ levels. We show that these deficiencies can be remedied by adding all local Feynman diagrams for the Cu-$d$ orbitals using the DMFT and treating the other states within GW approximation.

In the low-energy spectrum of La$_2$CuO$_4$, LSDA does not have an insulating solution; there is a single non-magnetic solution with zero energy gap as shown in the bandstructure(Fig. 3(a)) and total density of states (Fig. 4(a)). 
The non-spin-polarized MQSGW also predicts metal as shown in Fig. 4(a), but the
two bands of primarily $Cu\mbox{-}d_{x^2\mbox{-}y^2}$ character near the Fermi level
are well-separated from the rest of the bands (dashed lines in Fig. 3(b)). Spin-polarized MQSGW calculation (dashed lines in Fig. 3(c)) yields
qualitative different results from LSDA and non-spin-polarized MQSGW
calculation. The two $Cu\mbox{-}d_{x^2\mbox{-}y^2}$ bands are now well separated
from each other with a bandgap of 3.4 eV. Spin-polarized QSGW \cite{kotani_quasiparticle_2007} also yields insulating phase with a gap of 4.0 eV. In the experiment, the larger direct gap,
as measured by optics, is $\sim2eV$ ~\cite{ginder_photoexcitations_1988,
  cooper_optical_1990}.

We show that these deficiencies of LDA, QSGW and MQSGW in the low-energy spectra can be remedied by adding all
local Feynman diagrams for the Cu-$d$ orbitals using the
DMFT. 
The LDA+DMFT calculation in Fig. 4(a), carried out by
the all-electron LDA+DMFT
method~\cite{haule_dynamical_2010,haule_covalency_2014}, predicts reasonable gap of 1.5 eV and 1.8 eV in PM and AFM phases, in good agreement with
experiment and previous LDA+DMFT studies ~\cite{weber_optical_2008,weber_strength_2010,wang_covalency_2012,haule_covalency_2014,werner_dynamical_2014}. Within MQSGW+DMFT, we find gaps of 1.5 eV and 1.6 eV in PM and AFM phases, respectively, as shown in Fig. 4(b). The excitation spectra of MQSGW+DMFT in PM and AFM phase as shown in Fig. 3(b) and 3(c) are very similar as both are insulating with well separated
$Cu\mbox{-}d_{x^2\mbox{-}y^2}$ bands, which is now also substantially broadened due
to large scattering rate in Hubbard-like bands. In addition, MQSGW+DMFT improves the line-shape of LDA+DMFT. Near the top of the valence bands with oxygen $p$ character, the lineshape within LDA+DMFT is too sharp in comparison to the experiments as shown in Fig. 4(c). By treating oxygen $p$ levels within GW, the lineshape becomes smoother and in a better agreement with experiments. 

% The projected density of states of Cu-$d_{x^2\mbox{-}y^2}$
% orbital in Fig. 4(b) shows the size of the gap more apparently.  This gap is of correlated type as it results from the non-perturbative
% pole in the impurity self-energy near zero frequency. This pole is connected to spin fluctuations, which at low temperature results in an
% ordered AFM phase. The Zhang-Rice peak appears around $\sim$$-2eV$ and strongly overlapps with the onset of O-$p$ states.

In the high energy region of La$_2$CuO$_4$, the most distinctive difference is the position of
La-$f$ peak. It appears at $\sim3eV$ within LDA and LDA+DMFT, but at around $\sim 9eV$, in the inverse-photoemission spectra (cyan dotted line in Fig. 4(a)) \cite{nucker_experimental_1987}. By treating La-$f$ within GW approximation, it appears at $\sim10eV$ within MQSGW and MQSGW+DMFT.  
The underestimation of unoccupied La-$f$ excitation energy is attributed to the local approximation to the electron self-energy within LDA. Within LDA, Hartree and exchange-correlation potential applied to La-$f$ orbitals are orbital-independent since charge density is averaged over 14 different m channels \cite{anisimov_band_1991}. In contrast, these potentials within MQSGW are orbital-dependent and non-local. The effect of orbital-dependent potential can be tested within LDA+U approaches, since LDA+U adds orbital-dependent potential and subtracts orbital-independent potential explicitly \cite{anisimov_first-principles_1997}. From LDA+U approaches, we can also understand MQSGW better since LDA+U can be regarded as a local and static approximation to GW approximation \cite{anisimov_first-principles_1997}. According to M.T.Czyzyk and G.A.Sawatzky \cite{czyzyk_local-density_1994}, La-$f$ peaks shift from E$_F$+3eV to E$_F$+3eV+U/2 with U=11eV for La-$f$. %Also, the oxygen $p$ level (the main peak below the Fermi level) is shifted downward by 1 eV in MQSGW+DMFT relative to LDA+DMFT. This relative shift of $p$ bands within GW is well-known and its well established that it improves upon LDA \cite{aryasetiawan_gw_1998,faleev_all-electron_2004}.

\begin{figure}[t]
  \centering
  \includegraphics[width=0.70\columnwidth]{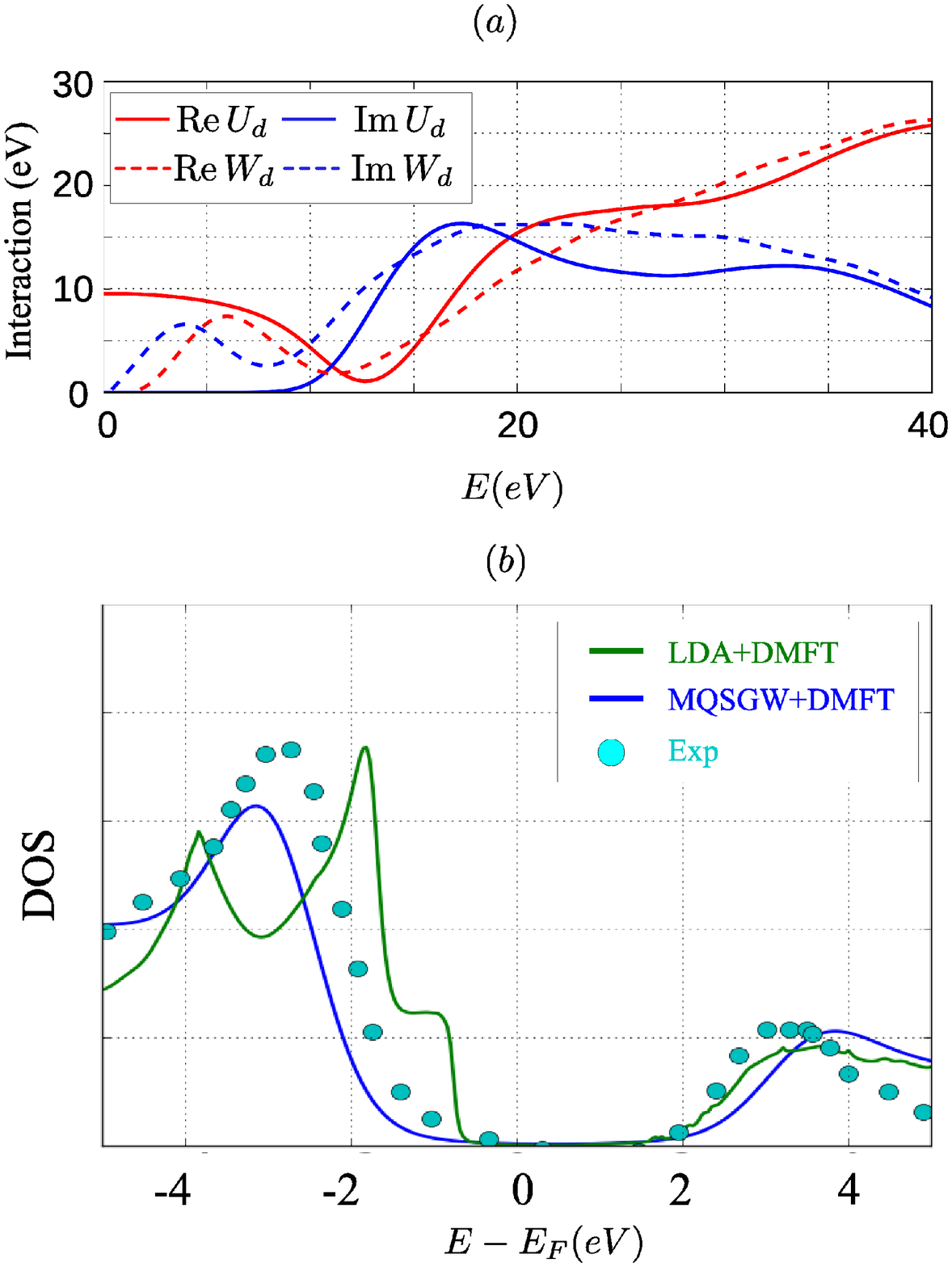}
  \caption {Hubbard U associated with Ni-3$d$ orbitals in NiO (a) Frequency dependence of $W_d$ (dashed lines) and $U_d$ (full lines) of NiO, with a $\chi_{QP}^{low}$ defined in the energy window in $E_F-11eV$ to $E_F+10eV$. Real and imaginary parts of the parameter are marked by red and blue colors, respectively. (b) Total density of states of NiO within LDA+DMFT(green) and MQSGW+DMFT(blue). The cyan dotted line shows photoemission/inverse-photoemission data \cite{zimmermann_electronic_1999}}
\end{figure}

We also tested our proposed scheme with one more charge transfer insulator, NiO. Fig. 5(a) shows the frequency dependence of $U_d$ and $W_d$ for the Ni-$3d$ orbitals in the low-energy region. In contrast to $W_d$, $U_d$ is almost constant up to $5\,$eV. Static $U_d$ is 9.6 eV. In the high energy limit, $U_d$ and $W_d$ approach the bare value of $26.0\,$eV. Calculated $J_H$ for the Ni-$3d$ orbitals has negligible frequency dependence and static $J_H$ is $1.4\,$eV. Fig. 5(b) shows the total density of states of NiO within LDA+DMFT and MQSGW+DMFT in its paramagnetic phase. Photoemission/inverse photoemission data are also plotted for comparison \cite{zimmermann_electronic_1999}. The LDA+DMFT calculation is being carried out by the all-electron LDA+DMFT method~\cite{haule_dynamical_2010} with $U_d=10eV, J_H=0.9eV$ and nominal double-counting energy. In the paramagnetic phase, LDA+DMFT and MQSGW+DMFT predict insulator in an agreement with previous LDA+DMFT studies \cite{ren_lda_2006,yin_calculated_2008}, but MQSGW+DMFT improves the line-shape of LDA+DMFT. Near the top of the valence bands, the lineshape within LDA+DMFT is too sharp in comparison to the experiments. By treating oxygen $p$ levels within GW, the lineshape becomes smoother and in a better agreement with experiments. In the antiferromagnetic phase, magnetic moment associated with Ni-$d$ orbitals is 1.6 $\mu_B$ within MQSGW+DMFT, in agreement with experimental value of 1.6-1.9 $\mu_B$ \cite{cheetham_magnetic_1983, fender_covalency_1968,ren_lda_2006}. 

In summary, we introduced a new methodology within MQSGW+DMFT and tested it in the classic charge
transfer insulator La$_2$CuO$_4$ and NiO. Our methodology predicts a
Mott-insulating gap in the PM phase, thus overcoming the limitation of
LDA and QSGW. It yields more precise peak positions of the La-$f$
states in La$_2$CuO$_4$ and valence band lineshape, thus improving the results of LDA+DMFT. The method should be
useful in understanding electronic excitation spectrum of other
strongly-correlated materials, in particular, those where precise
position of both the itinerant and correlated states is important.

\textbf{Methods}

 Our approach is carried it out entirely on the Matsubara axis, which requires
a different approach to the quasiparticle self-consistency in GW \cite{kutepov_electronic_2012}, called
Matsubara Quasiparticle Self-consistent GW (MQSGW), 
where  the  quasiparticle
Hamiltonian  is constructed by  linearizing the  self-energy and renormalization factor \cite{qsgw_vs_qpgw}.
Working on the Matsubara axis is numerically  very stable, provide a natural interface with advanced DMFT solvers such
as continuous-time quantum Monte-Carlo (CTQMC) \cite{werner_continuous-time_2006,haule_quantum_2007} and has very good scaling in system size as in the space-time method. 
(see supplementary note on Matsubara QSGW calculations).

For DMFT, it is essential to
obtain bandstructures in a fine enough crystal momentum ($\mathbf{k}$) mesh to attain desired
frequency resolution of physical quantities. To achieve such momentum
resolution, we use a Wannier-interpolated MQSGW bandstructure in a large
energy window using Maximally localized Wannier function (MLWF) \cite{marzari_maximally_2012}, and then constructed local projector in a
fine momentum mesh.
In contrast to SrVO$_3$
\cite{tomczak_combined_2012,sakuma_electronic_2013,taranto_comparing_2013,tomczak_asymmetry_2014}
where a set of $t_{2g}$ states is reasonably well separated from the
other bands, correlated $3d$ orbitals in La$_2$CuO$_4$ and NiO and are strongly
hybridized with other itinerant bands. In this case, it is necessary
to construct local projectors from states in a wide enough energy
windows to make projectors localized near the correlated atoms. We
constructed local projectors in the energy window $E_F\pm 10eV$ in which there are ${\sim}82$ bands at each $\mathbf{k}$
point, where $E_F$ is the Fermi level for La$_2$CuO$_4$. For NiO, we constructed local projectors in the energy window of $E_F-11eV$ to $E_F+10eV$. Then we confirmed that absolute value of
its overlap to the muffin-tin orbital (of which radial function is determined to maximize electron occupation in it) is more than 95\%. Our choice of energy
window is justified by the Cu-$3d$ spectra being entirely contained in this
window. Using constructed MLWFs, we defined our local-projector
$P_{i,n}(\mathbf{k})=\sum_{R}\left<W_{\mathbf{R}i}|\psi_{n\mathbf{k}}\right>e^{-i\mathbf{k}\cdot\mathbf{R}}/\sqrt{N_k}$,
where $W_{\mathbf{R}i}(\mathbf{r})$ is MLWF with an index $i$, $\psi_{n\mathbf{k}}(\mathbf{r})$ is quasiparticle wavefunction with
an index $n$, and $N_k$ is the number of $\mathbf{k}$ points in the first Brillouin zone. 

Static $U_d$ and $J_H$ are evaluated by a modification of the constrained
RPA method \cite{aryasetiawan_frequency-dependent_2004}, which avoids
screening by the strongly hybridized bands. This screening by
hybridization is included in our large energy window DMFT. For details, see supplementary note on $U_d$ and $J_H$. We divide
dynamic polarizability within MQSGW approximation $\chi_{QP}$ into two parts,
$\chi_{QP}=\chi_{QP}^{low}+\chi_{QP}^{high}$. Here,
$\chi_{QP}^{low}$ is defined by all transitions between the
states in the energy window accounted for by the DMFT method ($E_F\pm 10eV$ for La$_2$CuO$_4$ and $E_F-11eV$ to $E_F+10eV$ for NiO). 
Using $\chi_{QP}^{high}$, we evaluate partially-screened Coulomb interaction $U^{-1}(\mathbf{r},\mathbf{r}',\mathbf{k},i\omega_n)=V^{-1}(\mathbf{r},\mathbf{r}',\mathbf{k})-\chi_{QP}^{high}(\mathbf{r},\mathbf{r}',\mathbf{k},i\omega_n)$
and parametrize static $U_d$ and $J_H$ by Slater's
integrals \cite{van_der_marel_electron-electron_1988,kutepov_self-consistent_2010}, where $V$ is bare Coulomb interaction.

The Feynman graphs included  in both MQSGW and DMFT
(double-counting) are the local  Hartree and the local GW  diagram. They are computed
using  the local projection of the  MQSGW Green's function
($\hat{G}_{QP}$) $\hat{G}_{QP}^{loc}(i\omega_n)=\frac{1}{N_k}\sum_{\mathbf{k}}\hat{P}(\mathbf{k})\hat{G}_{QP}(\mathbf{k},i\omega_n)\hat{P}^\dagger(\mathbf{k})$
and the local Coulomb matrix constructed from Slater's integrals. For the details, see supplementary note on double counting energy.

\textbf{Acknowledgments}
This work was supported by the U.S Department of energy, Office of Science, Basic Energy Sicences as a part of the Computational Materials Science Program and by the Simons Foundation under  the Many Electron Problem  collaboration.  An award of computer time was provided by the INCITE program. This research used resources of the Oak Ridge Leadership Computing Facility, which is a DOE Office of Science User Facility supported under Contract DE-AC05-00OR22725.

\textbf{Author contributions}

G.K. designed the framework of the code. S.C developed the code, building on earlier developments of A.K. and K.H.  and performed the calculations. G.K., K.H. and S.C. analyzed the data with the help of A.K. and M.v.S. All authors provided comments on the paper.
% \bibliography{manuscript.bib}

\begin{thebibliography}{10}
\bibitem{optimal_u}
 \bibinfo{note}{The renormalization of $U_d$ by the nonlocal Coulomb
  interaction can be understood by using the definition of Hubbard U by C
  Herring. According to Herring\cite{rado1966magnetism}, the Hubbard U can be
  obtained from the reaction of $2(d^{n_d}p^{n_p})\to
  d^{n_d+1}p^{n_p-1}+d^{n_d-1}p^{n_p+1}$ between Cu-$d$ and O-$p$ orbitals,
  where $n_d$ and $n_p$ are the number of electrons in Cu-$d$ and O-$p$
  orbitals respectively. The energy cost for the reaction is approximately
  $U_{d}-2U_{pd}+U_{p}$ and this should be the sum of the renormalized $U_{d}$
  and renormalized $U_{p}$. In the simplest case, where we have one electron in
  $p$ and $d$ orbitals, renormalized U of Cu-d orbitals is $U_{d}-U_{pd}$ from
  the reaction of $d^{1}p^{1}\to d^{2}p^{0}$. The renormalized $U_d$ is derived
  more rigorously by sch{\''u}ler et al.\cite{schuler_optimal_2013}}.
\end{thebibliography}

\end{document}

\mathchardef\mhyphen="2D
\titlespacing{\section}{0pt}{2pt}{1ex}

\newcommand{\beginsupplement}{%
  \setcounter{table}{0}
  \renewcommand{\thetable}{S\arabic{table}}%
  \setcounter{figure}{0}
  \renewcommand{\thefigure}{S\arabic{figure}}%
  \setcounter{equation}{0}
  \renewcommand{\theequation}{S\arabic{equation}}
}
\beginsupplement
\title{Supplementary information for \emph{\textit{First-principles} treatment of Mott insulators: linearized QSGW+DMFT approach}}
\author{Sangkook Choi}
% \email[]{Your e-mail address}
% \homepage[]{Your web page}
% \thanks{}
% \altaffiliation{}
\affiliation{DMFT-MatDeLab center, Upton, New York 11973, USA}
\affiliation{Department of Physics and Astronomy, Rutgers University, Piscataway, New Jersey 08854, USA}
\author{Andrey Kutepov}
% \email[]{Your e-mail address}
% \homepage[]{Your web page}
% \thanks{}
% \altaffiliation{}
\affiliation{Department of Physics and Astronomy, Rutgers University, Piscataway, New Jersey 08854, USA}
\author{Kristjan Haule}
% \email[]{Your e-mail address}
% \homepage[]{Your web page}
% \thanks{}
% \altaffiliation{}
\affiliation{DMFT-MatDeLab center, Upton, New York 11973, USA}
\affiliation{Department of Physics and Astronomy, Rutgers University, Piscataway, New Jersey 08854, USA}
\author{Mark van Schilfgaarde}
% \email[]{Your e-mail address}
% \homepage[]{Your web page}
% \thanks{}
% \altaffiliation{}
\affiliation{Department of Physics, Kings College London, Strand, London WC2R 2LS, United Kingdom}
\author{Gabriel Kotliar}
% \email[]{Your e-mail address}
% \homepage[]{Your web page}
% \thanks{}
% \altaffiliation{}
\affiliation{DMFT-MatDeLab center, Upton, New York 11973, USA}
\affiliation{Department of Physics and Astronomy, Rutgers University, Piscataway, New Jersey 08854, USA}

\maketitle

\section*{Computational details: basis set}
Calculations are performed using our relativistic spin-polarized, full-potential, linearized-augmented-plane-wave package (RSPFLAPW) ~\cite{kutepov_ab_2003,kutepov_electronic_2012}, which is based on full-potential linearized augmented plane wave plus local orbital method. For La$_2$CuO$_4$, experimental lattice constants and atomic positions at the low-temperature orthorhombic phase \cite{reehuis_crystal_2006} are used. The following parameters for the basis are used: muffin-tin (MT) radii ($R_{MT}$) in Bohr radius are 2.2 for Lanthanum(La), 2.1 for Cu, and 1.4 for Oxygen(O). Wave functions are expanded by spherical harmonics with $l$ up to 4 for La, 4 for Cu, and 3 for O in the MT spheres, and by plane waves with the energy cutoff determined by $R_{MT,Cu} \times K_{max} =7.8$ in the interstitial (IS) region. The Brillioun zone was sampled with $3 \times 3 \times 3 $ $\mathbf{k}$-point grid. For the density functional theory calculation, local spin density approximation (LSDA) is employed. For the GW calculation, The following convergence parameters are used. Product basis is expanded by spherical harmonics with $l$ up to $l_{max}$ =5 in the MT spheres and $R_{MT,Cu} \times K_{max} =9.0$ in IS region. Unoccupied states with an energy up to 500 eV from Fermi energy are taken into account for polarizability and self-energy calculation. The projector is constructed using MLWF and interpolated in a $10\times10\times10$ $\mathbf{k}$-grid. Spin-orbit coupling was not included.

For NiO in the rocksalt structure with a lattice constant  of 4.19 $\AA$ \cite{pearson_handbook_1967}, the following parameters for the basis are used: muffin-tin (MT) radii ($R_{MT}$) in Bohr radius are 2.12 for Nickel(Ni) and 1.77 for O. Wave functions are expanded by spherical harmonics with $l$ up to 6 for Ni and 6 for O in the MT spheres, and by plane waves with the energy cutoff determined by $R_{MT,Ni} \times K_{max} =6.7$ in the interstitial (IS) region The Brillioun zone was sampled with $5 \times 5 \times 5 $ $\mathbf{k}$-point grid. For the GW calculation, The following convergence parameters are used. Product basis are expanded by spherical harmonics with $l$ up to $l_{max}$ =6 in the MT spheres and $R_{MT,Cu} \times K_{max} =10.0$ in IS region. Unoccupied states with an energy up to 680 eV from Fermi energy are taken into account for polarizability and self-energy calculation. The projector is constructed using MLWF and interpolated in a $15\times15\times15$ $\mathbf{k}$-grid. Spin-orbit coupling was not included.
\\
\section*{Matsubara QSGW calculations}
The electron self-energy can be systematically expanded in terms of
the dressed Green's function $G$ and the
screened Coulomb interaction $W$. Within GW approximation, we keep only
the first term of the series expansion of self-energy in $W$:
\begin{equation}
  \begin{split} \Sigma_{GW}(\mathbf{r},\mathbf{r}',\mathbf{k},s,i\omega_n)=-\sum_{\mathbf{R}}\int{d\tau}G(\mathbf{r},\mathbf{r}',\mathbf{R},s,\tau)W(\mathbf{r},\mathbf{r}',\mathbf{R},\tau)e^{-i(\mathbf{k}\cdot\mathbf{R}-\omega_n\tau)}, 
  \end{split}
\end{equation}
where $\mathbf{r}$ is the position vector in a unit cell, $\mathbf{k}$ is the
crystal momentum, $\mathbf{R}$ is the
lattice vector, $\omega_n$ is Matsubara frequency, $\tau$ is Matsubara time, and $s$ is spin index. Within Matsubara quasiparticle self-consistent GW (MQSGW) approximation, we calculate the dynamic polarizability and the electron self-energy by using quasiparticle Green's function ($G_{QP}$) instead of Full GW Green's function. First, we construct quasiparticle (QP) Green's function using the quasiparticle Hamiltonian ($H_{QP}$) by 
\begin{equation}
  \begin{split}
    G_{QP}^{-1}(\mathbf{r},\mathbf{r}',\mathbf{k},i\omega_n)=i\omega_n\delta(\mathbf{r}-\mathbf{r}')-H_{QP}(\mathbf{r},\mathbf{r}',\mathbf{k}).
  \end{split}
\end{equation}
For the first iteration, we regard Hamiltonian within local density approximation as $H_{QP}$. Next, the screened Coulomb interaction $W_{QP}$ is evaluated using the dynamic polarizability $\chi_{QP}(\mathbf{r},\mathbf{r}',\mathbf{k},'i\omega_n)$ by $W_{QP}^{-1}(\mathbf{r},\mathbf{r}',\mathbf{k},i\omega_n)=V^{-1}(\mathbf{r},\mathbf{r}',\mathbf{k})-\chi_{QP}(\mathbf{r},\mathbf{r}',\mathbf{k},i\omega_n)$
within random phase approximation (RPA): 
\begin{equation}
  \begin{split}
    \chi_{QP}(\mathbf{r},\mathbf{r}',\mathbf{k},'i\omega_n)=\sum_s\sum_\mathbf{R}\int d\tau G_{QP}(\mathbf{r},\mathbf{r}',\mathbf{R},s,\tau)G_{QP}(\mathbf{r}',\mathbf{r},-\mathbf{R},s,-\tau) e^{-i(\mathbf{k}\cdot\mathbf{R}-\omega_n\tau)},
  \end{split}
\end{equation}
where $V$ is bare Coulomb interaction. Then, we calculate MQSGW electron self-energy ($\Sigma_{QP}$) by
\begin{equation}
  \begin{split} \Sigma_{QP}(\mathbf{r},\mathbf{r}',\mathbf{k},s,i\omega_n)=-\sum_{\mathbf{R}}\int{d\tau}G_{QP}(\mathbf{r},\mathbf{r}',\mathbf{R},s,\tau)W_{QP}(\mathbf{r},\mathbf{r}',\mathbf{R},\tau)e^{-i(\mathbf{k}\cdot\mathbf{R}-\omega_n\tau)}.
  \end{split}
\end{equation}
Then, we constructed quasiparticle Hamiltonian with linearized self-energy and renormalization factor, $Z^{-1}(\mathbf{r},\mathbf{r}',\mathbf{k})= 1-\partial\Sigma_{QP}(\mathbf{r},\mathbf{r}',\mathbf{k},i\omega_n=0)/\partial (i\omega_n)$ by
\begin{equation}
  \begin{split}    \hat{H}_{QP}(\mathbf{k})=\hat{Z}^{1/2}(\mathbf{k})\left(\hat{H}_{H}(\mathbf{k})+\hat{\Sigma}_{QP}(\mathbf{k},i\omega_n=0)\right)\hat{Z}^{1/2}(\mathbf{k}),
  \end{split}
\end{equation}
where $\hat{H}_{H}(\mathbf{k})$ is Hartree Hamiltonian. This process is repeated until self-consistency is attained.
\\
\section*{$U_d$ and $J_H$}
The static $U_d$ and $J_H$ are evaluated by a method similar to constrained
RPA~\cite{aryasetiawan_frequency-dependent_2004}, but here we avoid
screening by the strongly hybridized bands, which is included in our large energy window DMFT.  We divide
dynamic polarizability $\chi_{QP}$ into two parts,
$\chi_{QP}=\chi_{QP}^{low}+\chi_{QP}^{high}$. Here,
$\chi_{QP}^{low}$ is defined by all transitions between the
states in the energy window accounted for by the DMFT method ($E_F\pm 10eV$ for La$_2$CuO$_4$ and $E_F-11eV$ to $E_F+10eV$ for NiO):
\begin{equation}
  \begin{split}
    \chi_{QP}^{low}(\mathbf{r},\mathbf{r}',\mathbf{k},i\omega_n)=-2\sum_{\mathbf{k'}}\sum_{n}^{\substack{\text{unocc}\\\text{in the window}}}\sum_{m}^{\substack{\text{occ}\\\text{in the window}}}\\
    \psi_{n\mathbf{k}'}(\mathbf{r})\psi_{m\mathbf{k}'+\mathbf{k}}^*(\mathbf{r})\psi_{n\mathbf{k}'}^*(\mathbf{r}')\psi_{m\mathbf{k}'+\mathbf{k}}(\mathbf{r}')\frac{2(E_{n\mathbf{k}'}-E_{n\mathbf{k}'+\mathbf{k}})}{\omega_n^2-(E_{n\mathbf{k}'}-E_{n\mathbf{k}'+\mathbf{k}})^2}
  \end{split}
\end{equation}
Using $\chi_{QP}^{high}$, we evaluate partially-screened Coulomb interaction by
$U_{}^{-1}(\mathbf{r},\mathbf{r}',\mathbf{k},i\omega_n)=V^{-1}(\mathbf{r},\mathbf{r}',\mathbf{k})-\chi_{QP}^{high}(\mathbf{r},\mathbf{r}',\mathbf{k},i\omega_n)$.
Then, we calculate $U=F_0$ and $J_H=(F_2+F_4)/14$ by using Slater
integrals \cite{kutepov_self-consistent_2010},
\begin{equation}
  \begin{split}
    F_k=\frac{1}{C_{k}}\frac{4\pi}{2k+1}\sum_{m_1,m_2,m_3,m_4}\bra{Y_{2m_1}}Y_{km_1-m_4}Y_{2m_4}\rangle\bra{Y_{2m_2}Y_{km_2-m_3}}Y_{2m_3}\rangle\\
    \int d\mathbf{r}d\mathbf{r}'U(\mathbf{r},\mathbf{r}', R=0,i\omega_n=0)W_{R=0,m_1}^*(\mathbf{r})W_{R=0,m_2}^*(\mathbf{r}')W_{R=0,m_3}(\mathbf{r}')W_{R=0,m_4}(\mathbf{r}),\\
  \end{split}
\end{equation}
where $W_{\mathbf{R},i}(\mathbf{r})$ is Wannier function centered at $\mathbf{R}$ with index $m$ and $Y_{lm}$ is spherical harmonics. Here, $C_0=25$, $C_2=20/49$ and $C_4=100/441$. 
\\

\section*{DMFT: local projection and embedding}
With non-spin-polarized MQSGW Hamiltonian ($H_{QP}$), we solve DMFT self-consistent equation, 
\begin{equation}
  \begin{split}
    &\frac{1}{N_k}\sum_{\mathbf{k}}P_{i,n_1}(\mathbf{k})\left\{i\omega_n\mathbf{1}-\hat{H}_{QP}(\mathbf{k},i\omega_n)-\hat{\Sigma}_{embed}(\mathbf{k},s,i\omega_n)\right\}^{-1}_{n_1,n_2}P_{j,n_2}^*(\mathbf{k})\\
    &={(i\omega_n\mathbf{1}-\hat{E}_{imp}-\hat{\Delta}_{imp}(s,i\omega)-\hat{\Sigma}_{imp}(s,i\omega))}_{i,j}^{-1}\\
  \end{split}
\end{equation}
Here, $P_{i,n}(\mathbf{k})=\sum_{R}<W_{\mathbf{R}i}|\psi_{n\mathbf{k}}>e^{i\mathbf{k}\cdot\mathbf{R}}/\sqrt{N_k}$ is projector to the correlated subspace at each $\mathbf{k}$ and $\psi_{n\mathbf{k}}(\mathbf{r})$ is quasiparticle wavefunction with an index $n$. $N_k$ is the number of $\mathbf{k}$ points in the first Brillouin zone. $\hat{E}_{imp}$ and $\hat{\Delta}_{imp}$ are impurity level energy and hybridization function, which are inputs to impurity solver. $\hat{\Sigma}_{embedded}=P^{\dagger}(\mathbf{k})\left(\hat{\Sigma}_{imp}(s,i\omega_n)-\hat{\Sigma}_{DC}(i\omega_n)\right)\hat{P}(\mathbf{k})$ is embedded self-energy with impurity self-energy ($\hat{\Sigma}_{imp}$) and double-counting correction ($\hat{\Sigma}_{DC}$). 
\\

\section*{Double counting energy}
The Feynman graphs included  in both MQSGW and DMFT
(double-counting) are the local  Hartree and the local GW  diagram. They are computed
using  the local projection of the  MQSGW Green's function
($\hat{G}_{QP}$) $\hat{G}_{QP}^{loc}(i\omega_n)=\frac{1}{N_k}\sum_{\mathbf{k}}\hat{P}(\mathbf{k})\hat{G}_{QP}(\mathbf{k},i\omega_n)\hat{P}^\dagger(\mathbf{k})$
and the local Coulomb matrix $U_{iklj}$:
\begin{equation}
  \Sigma_{i,j}^{DC}(i\omega_n)=\sum_{k,l=Cu \mhyphen d} 2G_{QP, l,k}^{loc}(\tau=0^-)U_{iklj}-\sum_{k,l=Cu\mhyphen d}\int d\tau G_{QP,l,k}^{loc}(\tau)W_{ikjl}^{loc}(\tau)e^{i \omega_n\tau},\label{eq:dc}
\end{equation}
where, $W_{ikjl}^{loc}(i\omega_n){=}U_{ikjl}+\sum_{mnpq=Cu\mhyphen
  d}\allowbreak U_{imnl} \allowbreak\chi_{mpqn}^{loc}(i\omega_n)\allowbreak W_{pkjq}^{loc}(i\omega_n)$
and $\chi_{mpqn}^{loc}(i\omega_n)\allowbreak=2\int d\tau$
$G_{QP,n,p}^{loc}(\tau)G_{QP,q,m}^{loc}(-\tau)\allowbreak e^{i\omega\tau}$.
The Coulomb matrix $U_{iklj}$ is constructed by using Slater integrals of $F_0=12.0eV$, $F_2=12.1eV$, and $F_4=7.5eV$ for La$_2$CuO$_4$ and $F_0=9.6eV$, $F_2=12.1eV$, and $F_4=7.5eV$ for NiO in the following way:
\begin{equation}
  \begin{split}
    U^{m1,m2,m3,m4}(i\nu_n)&=S_{m_1,m_1'}S_{m_2,m_2'}S_{m_3,m_3'}^{-1}S_{m_4,m_4'}^{-1}\\
&\sum_{k=0}^{2l}\frac{4\pi}{2k+1}\bra{Y_{2m_1'}}Y_{kq}Y_{2m_4'}\rangle\bra{Y_{2m_2'}Y_{kq}}Y_{2m_3'}\rangle F^{k}(i\nu_n)
    \label{eq:u_from_f}
  \end{split}
\end{equation}
where $S$ is the transformation matrix from spherical harmonics to cubic harmonics.
Finally, for the stable numerics, we approximated
$\hat{\Sigma}^{DC}(i\omega_n)\simeq\hat{\Sigma}^{DC}(i\omega_n=0)$
since these low order diagrams are dominated by the Hartree-Fock contribution.
% \allowbreak{=}\allowbreak\int d\mathbf{r} d\mathbf{r}' \allowbreak W_{\mathbf{R}{=}0,i}^*(\mathbf{r}) W_{\mathbf{R}{=}0,k}^*(\mathbf{r}') \allowbreak  W_{\mathbf{R}{=}0,l}(\mathbf{r}')\allowbreak  W_{\mathbf{R}{=}0,j}(\mathbf{r})\allowbreak  U(\mathbf{r},\mathbf{r}',\mathbf{R}{=}0,i\omega_n{=}0)$:
\\